\documentclass[5p,11pt,fleqn,sort&compress,times]{elsarticle}

\usepackage{amsmath,amssymb,bbm,multirow,graphicx}
\usepackage[colorlinks,citecolor=blue,urlcolor=blue,linkcolor=blue]{hyperref}

\DeclareMathOperator{\diag}{diag}
\DeclareMathOperator{\im}{Im}
\DeclareMathOperator{\Tr}{Tr}

\journal{Phys. Lett. B}

\begin{document}

\begin{frontmatter}

\title{More about unphysical zeroes in quark mass matrices} 

\author[ist]{David Emmanuel-Costa}
\ead{david.costa@tecnico.ulisboa.pt}

\author[isel,ist]{Ricardo Gonz\'{a}lez Felipe}
\ead{ricardo.felipe@tecnico.ulisboa.pt}

\address[ist]{Departamento de F\'{\i}sica and Centro de F\'{\i}sica Te\'{o}rica de Part\'{\i}culas - CFTP, Instituto Superior T\'{e}cnico, Universidade de Lisboa,\\ Avenida Rovisco Pais, 1049-001 Lisboa, Portugal}
	
\address[isel]{ISEL - Instituto Superior de Engenharia de Lisboa, Instituto Polit\'ecnico de Lisboa,\\ Rua Conselheiro Em\'{\i}dio Navarro 1959-007 Lisboa, Portugal}

\begin{abstract}
We look for all weak bases that lead to texture zeroes in the quark mass matrices and contain a minimal number of parameters in the framework of the standard model. Since there are ten physical observables, namely, six nonvanishing quark masses, three mixing angles and one CP phase, the maximum number of texture zeroes in both quark sectors is altogether nine. The nine zero entries can only be distributed between the up- and down-quark sectors in matrix pairs with six and three texture zeroes or five and four texture zeroes. In the weak basis where a quark mass matrix is nonsingular and has six zeroes in one sector, we find that there are 54 matrices with three zeroes in the other sector, obtainable through right-handed weak basis transformations. It is also found that all pairs composed of a nonsingular matrix with five zeroes and a nonsingular and nondecoupled matrix with four zeroes simply correspond to a weak basis choice. Without any further assumptions, none of these pairs of up- and down-quark mass matrices has physical content. It is shown that all non-weak-basis pairs of quark mass matrices that contain nine zeroes are not compatible with current experimental data. The particular case of the so-called nearest-neighbour-interaction pattern is also discussed. 
\end{abstract}

\begin{keyword}
Quark masses and mixing \sep Weak basis transformations 
\PACS 12.15Ff \sep 12.15Hh \sep 14.65.-q\\
\emph{Preprint:} CFTP/16-012\\
\emph{arXiv:} 1609.09491 [hep-ph]
\end{keyword}

\end{frontmatter}

\section{Introduction}
\label{sec:intro}

In the absence of a convincing theory to resolve the flavour puzzle in the standard model (SM), several approaches have been investigated. Among them, the systematic search for texture zeros in the mass matrices~\cite{Ramond:1993kv} has been quite popular, since such patterns usually contain less free parameters and, in some cases, they lead to relations between the flavour mixing angles and mass ratios that could be testable. From the theoretical point of view,  such texture zeroes have also been motivated by the introduction of symmetries in the Lagrangian~\cite{Grimus:2004hf,Felipe:2014vka}.

A common difficulty in the search for experimentally viable quark mass matrices with texture zeros is the fact that some of these zeroes may not have physical content. Indeed,  starting from arbitrary fermion mass matrices, certain matrix elements can be set to zero by making appropriate  transformations that leave the gauge currents flavour diagonal and real~\cite{Branco:1999nb,Falcone:1999mf,Branco:2007nn,EmmanuelCosta:2009bx,Ludl:2015lta} - the so-called weak basis (WB) transformations. These zeroes are usually referred to as WB zeroes. Two sets of quark mass matrices related by a WB transformation obviously have the same physical content.

In view of the freedom in the WB choice, it is then important to distinguish between physical and unphysical zeroes. In the SM, the flavour structure of  the Yukawa sector is not constrained by the gauge symmetry.  The quark mass matrices are $3\times3$ arbitrary complex matrices with altogether 36 free real parameters. This number is to be compared to the ten physical parameters corresponding to six nonvanishing quark masses plus three mixing angles and one CP phase in the standard parametrisation of the Cabibbo-Kobayashi-Maskawa (CKM) matrix.

In this work, we will analyse systematically the physical content of quark texture zeros within the SM framework. We look for all weak bases that lead to texture zeroes in the quark mass matrices and contain the maximal number of vanishing matrix elements. In Section~\ref{sec:gen}, we revise some general concepts related to WB transformations. It is shown that a pair of quark mass matrices with more than nine zeroes (distributed between them) necessarily contains less than 10 independent real parameters and, therefore, the maximum number of texture zeroes that could be obtained through WB transformations in both quark sectors is altogether nine. There are only two ways to distribute these zeroes between the up- and down-quark mass matrices, namely, in pairs with six and three texture zeroes or with five and four texture zeroes.\footnote{Obviously, more than six zeroes in a quark sector would lead to massless quarks.} We consider in Section~\ref{sec:RhWBT} all possible right-handed rotations in order to obtain the maximal number of zeroes that can be set in a given quark mass matrix. Starting from the WB in which one quark sector is diagonal, we show that it is always possible to construct, by means of right-handed rotations, quark mass matrices in the other sector that have three zeroes. In Section~\ref{sec:genWB}, we then look for WB left-handed rotations, which, after being applied simultaneously in both quark sectors, and combined with additional right-handed rotations and permutations, lead to pairs of quark mass matrices with five and four WB texture zeroes. Clearly, none of these pairs of up- and down-quark mass matrices has physical content. The particular case of the so-called nearest-neighbour-interaction (NNI) pattern is also discussed in this context. Quark mass matrices that contain the maximal number of texture zeroes, without being a WB choice, are confronted with current experimental data in Section~\ref{sec:numex}. Finally, our conclusions are given in Section~\ref{sec:summary}.

\section{General concepts}
\label{sec:gen}

In the context of the SM, the quark mass matrices $M_u$ and $M_d$ encode a large redundancy. This can be understood by the freedom one has to perform WB transformations in the up-quark fields $u_{L,R}$ and down-quark fields $d_{L,R}$:
\begin{subequations}
\label{eq:WBT0}
\begin{align}
u_L\,&=\,W_L\,\widetilde{u}_L,\qquad u_R\,=\,W_R^u\,\widetilde{u}_R,\\
d_L\,&=\,W_L\,\widetilde{d}_L,\qquad d_R\,=\,W_R^d\,\widetilde{d}_R,
\end{align}
\end{subequations}
where $W_L$, $W_R^u$ and $W_R^d$ are arbitrary unitary matrices. With such transformations, the gauge currents remain flavour diagonal and real, but the matrices $M_u\,\rightarrow \widetilde{M}_u$ and $M_d\,\rightarrow \widetilde{M}_d$ change as follows:
\label{eq:WBTm}
\begin{equation}
 \widetilde{M}_u=\,W_L^{\dagger}\,M_u\,W^u_R,\quad
 \widetilde{M}_d=\,W_L^{\dagger}\,M_d\,W^d_R,
\end{equation}
without altering the physical content. Note that any WB transformation applied on the right  can be performed independently in each quark sector. On the other hand, a WB transformation acting on the left  must be simultaneously the same in both sectors.  In the following, a WB transformation that acts only on the right of a quark mass matrix (i.e., with $W_L=\mathbbm{1}$) is referred to as a right-handed WB transformation (RhWBT). Similarly, a WB transformation that acts only on the left of  both quark mass matrices $M_u$ and $M_d$ (i.e., with $W^u_R=W^d_R=\mathbbm{1}$) is referred to as a left-handed WB transformation (LhWBT).

Among many possible weak bases, the physical basis where one quark mass matrix is diagonal, while the other is Hermitian, is particularly economical. In this case, one can write
\begin{subequations}
\label{eq:physDu}
\begin{align}\
M_u&=\diag(m_u,\,m_c,\,m_t),\\
M_d&=V\diag(m_d,\,m_s,\,m_b)V^{\dagger},
\label{eq:physDu-b}
\end{align}
\end{subequations}
or
\begin{subequations}
\label{eq:physDd}
\begin{align}\
\label{eq:physDd-a}
M_u&=V^{\dagger}\diag(m_u,\,m_c,\,m_t)V,\\
M_d&=\diag(m_d,\,m_s,\,m_b),
\end{align}
\end{subequations}
in terms of the quark masses $m_q$, $q={u,d,s,c,b,t}$, and the unitary CKM mixing matrix $V$. 

There is still the freedom to rephase the non-diagonal mass matrix elements in Eq.~\eqref{eq:physDu-b} or~\eqref{eq:physDd-a} through unitary transformations of the type
\begin{equation}
M_a\rightarrow K^{\dagger}M_aK,
\end{equation}
where $a=u$ or $d$, and the matrix $K$ is unitary and diagonal.  Such transformations can also be seen as a rephasing of the CKM matrix $V$.  In this new WB,  $M_a$ remains Hermitian and  there is only one independent rephasing invariant phase, defined as
\begin{equation}
\varphi_a = \arg\left[{\left(M_a\right)}_{12}^{\phantom{\ast}}{\left(M_a\right)}_{23}^{\phantom{\ast}}{\left(M_a\right)}_{13}^{\ast}\right].
\end{equation}
This phase is directly related to the well-known CP-odd WB invariant that controls CP violation in the quark sector. Indeed, this invariant can be written as\footnote{The proportionality factor involves the product of quark mass differences and thus the invariant vanishes when the quark spectrum has degeneracy.}
\begin{equation}
\im\Tr\left[M_u,M_d\right]^3\propto\left|{\left(M_a\right)}_{12}{\left(M_a\right)}_{23}{\left(M_a\right)}^{\ast}_{13}\right|\sin\varphi_a.
\end{equation}
Note also that the imaginary part of the product of matrix elements ${\left(M_a\right)}_{12}^{\phantom{\ast}}{\left(M_a\right)}_{23}^{\phantom{\ast}}{\left(M_a\right)}_{13}^{\ast}$ can be expressed through the CP invariant as
\begin{multline}
\im\left[{\left(M_a\right)}_{12}{\left(M_a\right)}_{23}{\left(M_a\right)}_{13}^{\ast}\right] =\\
(m^a_3-m^a_1)(m^a_3-m^a_2)(m^a_2-m^a_1)\im\left[V^{\phantom{\ast}}_{12}V^{\phantom{\ast}}_{23}V_{22}^{\ast}V_{13}^{\ast}\right],
\end{multline}
where $m^a_{1,2,3}$ are the physical masses encoded in the mass matrix $M_a$.

We conclude that in the WB where one quark mass matrix is diagonal and the other one is Hermitian, there are only ten independent parameters, namely three quark masses from the diagonal sector plus six moduli $|({M_a})_{ij}|$ and one rephasing invariant phase $\varphi_a$ arising from the matrix $M_a$. This is the simplest WB in which the quark mass matrices $M_u$ and $M_d$ are directly written in terms of the 10 physical observables.

One may wonder whether there exist other weak bases as economical as the one previously discussed, i.e., with just ten independent real parameters. An obvious possibility is to relax the Hermiticity condition and impose the vanishing of some matrix elements (texture zeroes). In this case, the elements of one of the mass matrices, let us say $M$, can be rephased through  transformations of the type
$M\rightarrow K^{\dagger}MK'$, where $K$ and $K'$ are diagonal unitary matrices. Since these transformations leave 
invariant any quartet combination $M^{\phantom{\ast}}_{ij} M^{\phantom{\ast}}_{kl} M^{\ast}_{il} M^{\ast}_{kj}$, the above transformation does not allow to remove simultaneously all the phases of the matrix elements belonging to the same quartet. Thus, given a pair of arbitrary $3\times3$ complex matrices $M_u$ and $M_d$, we can always remove five phases from one matrix, provided that any  four of these phases do not belong to elements of a quartet. There is still freedom to rephase away three phases at any position in the second matrix.

Since the number of physical parameters is ten, from the above analysis we conclude that the maximum number of zeroes that a weak basis can have is $(36-8-10)/2=9$. Therefore, starting from a WB where one of quark mass matrices is diagonal and real, and the other one is an arbitrary complex matrix $M$, the maximal number of texture zeroes that could be obtained through WB transformations is three. Remarkably, the three zeroes can be achieved by means of RhWBTs, as shown in the next section. 

\begin{table}[t]
\caption{\label{tab:classes} Classification of textures zeroes in a $3\times3$ complex mass matrix $M$. In the table, $N_{dec}$ stands for the number of textures where a quark generation is decoupled.}
\begin{center}
\begin{tabular*}{\linewidth}{@{\extracolsep{\fill}}cccccc}
\hline
  zeroes 
 & 
  textures 
 & 
 $|M|=0$ & $N_{dec}$  & $H_{ij}\neq 0$ & $H_{ij}=0$ \\
 & & & & {\small for all $i,j$} & {\small for $i\neq j$}\\
\hline
1 & 9 & -- & -- & 9 & -- \\
2 & 36 & -- & -- & 36 & -- \\ 
3 & 84 & 6 & -- & 60 &18  \\
4 & 126 & 45 & 9 & 18 &54 \\
5 & 126 & 90 & 36 & -- &-- \\
6 & 84 & 78 & 6 & -- &-- \\
\hline
\end{tabular*}
\end{center}
\end{table}

In Table~\ref{tab:classes}, we enumerate all possible zero textures for a given quark mass matrix $M$ containing up to 6 zeroes. To conveniently classify these textures we make use of the Hermitian matrix $H\equiv\,M\,M^{\dagger}$, since this matrix is invariant under RhWBT. In the table, for each type of zero-texture, we present the number of textures with vanishing determinant, those textures having a decoupled quark generation (i.e., $H$ has two or three null off-diagonal elements), the number of textures for which $H$ has no zeroes, and, finally, textures where only one off-diagonal matrix element in $H$ vanishes ($H_{12}=0$, $H_{13}=0$, or $H_{23}=0$).

By now it should be clear that having 10 zeroes distributed in $M_u$ and $M_d$ does not correspond to a WB choice. In other words, such patterns are necessarily ans\"atze, since there are 9 real parameters in $M_u$ and $M_d$ to fit the 10 physical parameters. There are only two types of potentially viable textures with 10 zeroes, namely, five zeroes in each quark sector or four zeroes in one sector and six zeroes in the other. Having five zeroes in each quark mass matrix would lead to a decoupled generation in each quark sector and therefore no CP violation phenomena could be accommodated. In the case of four-zero mass matrices, two classes can be distinguished: a matrix in which one quartet persists with nonvanishing elements, implying a null column or row in the corresponding quark mass matrix, or a matrix where all phases can be rephased away, meaning that CP is conserved in the quark sector. 

One then concludes that nine zeroes is not only the maximum amount of zeroes attainable through WB transformations, but it is also the maximum number of viable texture zeroes for a given quark mass matrix pair $M_u$ and $M_d$. We emphasise that this conclusion is general, independently of the origin of the texture zeroes.    

\section{Right-handed weak basis transformations}
\label{sec:RhWBT}

\begin{table*}[t]
	\caption{\label{tab:3ZWB} The 54 matrices with three zeros that can be obtained through RhWBT.}
		\begin{tabular*}{\linewidth}{@{\extracolsep{\fill}}ccccccccc}
			\hline
			$\begin{pmatrix}
			0 & 0 & \mathsf{x} \\
			0 & \mathsf{x} & \mathsf{x} \\
			\mathsf{x} & \mathsf{x} & \mathsf{x} \\
			\end{pmatrix}$
			& $\begin{pmatrix}
			0 & 0 & \mathsf{x} \\
			\mathsf{x} & 0 & \mathsf{x} \\
			\mathsf{x} & \mathsf{x} & \mathsf{x} \\
			\end{pmatrix}$
			& $\begin{pmatrix}
			\mathsf{x} & 0 & 0 \\
			\mathsf{x} & \mathsf{x} & 0 \\
			\mathsf{x} & \mathsf{x} & \mathsf{x} \\
			\end{pmatrix}$
			& $\begin{pmatrix}
			0 & \mathsf{x} & 0 \\
			0 & \mathsf{x} & \mathsf{x} \\
			\mathsf{x} & \mathsf{x} & \mathsf{x} \\
			\end{pmatrix}$
			& $\begin{pmatrix}
			0 & \mathsf{x} & 0 \\
			\mathsf{x} & \mathsf{x} & 0 \\
			\mathsf{x} & \mathsf{x} & \mathsf{x} \\
			\end{pmatrix}$
			& $\begin{pmatrix}
			\mathsf{x} & 0 & 0 \\
			\mathsf{x} & 0 & \mathsf{x} \\
			\mathsf{x} & \mathsf{x} & \mathsf{x} \\
			\end{pmatrix}$
			&	
			$\begin{pmatrix}
			0 & \mathsf{x} & \mathsf{x} \\
			0 & 0 & \mathsf{x} \\
			\mathsf{x} & \mathsf{x} & \mathsf{x} \\
			\end{pmatrix}$
			& $\begin{pmatrix}
			\mathsf{x} & 0 & \mathsf{x} \\
			0 & 0 & \mathsf{x} \\
			\mathsf{x} & \mathsf{x} & \mathsf{x} \\
			\end{pmatrix}$
			& $\begin{pmatrix}
			\mathsf{x} & \mathsf{x} & 0 \\
			\mathsf{x} & 0 & 0 \\
			\mathsf{x} & \mathsf{x} & \mathsf{x} \\
			\end{pmatrix}$
			\\[5mm]
			 $\begin{pmatrix}
			0 & \mathsf{x} & \mathsf{x} \\
			0 & \mathsf{x} & 0 \\
			\mathsf{x} & \mathsf{x} & \mathsf{x} \\
			\end{pmatrix}$
			& $\begin{pmatrix}
			\mathsf{x} & \mathsf{x} & 0 \\
			0 & \mathsf{x} & 0 \\
			\mathsf{x} & \mathsf{x} & \mathsf{x} \\
			\end{pmatrix}$
			& $\begin{pmatrix}
			\mathsf{x} & 0 & \mathsf{x} \\
			\mathsf{x} & 0 & 0 \\
			\mathsf{x} & \mathsf{x} & \mathsf{x} \\
			\end{pmatrix}$
			&
			$\begin{pmatrix}
			0 & \mathsf{x} & \mathsf{x} \\
			0 & \mathsf{x} & \mathsf{x} \\
			\mathsf{x} & 0 & \mathsf{x} \\
			\end{pmatrix}$
			& $\begin{pmatrix}
			\mathsf{x} & 0 & \mathsf{x} \\
			\mathsf{x} & 0 & \mathsf{x} \\
			0 & \mathsf{x} & \mathsf{x} \\
			\end{pmatrix}$
			& $\begin{pmatrix}
			\mathsf{x} & \mathsf{x} & 0 \\
			\mathsf{x} & \mathsf{x} & 0 \\
			\mathsf{x} & 0 & \mathsf{x} \\
			\end{pmatrix}$
			&
			$\begin{pmatrix}
			0 & \mathsf{x} & \mathsf{x} \\
			0 & \mathsf{x} & \mathsf{x} \\
			\mathsf{x} & \mathsf{x} & 0 \\
			\end{pmatrix}$
			& $\begin{pmatrix}
			\mathsf{x} & \mathsf{x} & 0 \\
			\mathsf{x} & \mathsf{x} & 0 \\
			0 & \mathsf{x} & \mathsf{x} \\
			\end{pmatrix}$
			& $\begin{pmatrix}
			\mathsf{x} & 0 & \mathsf{x} \\
			\mathsf{x} & 0 & \mathsf{x} \\
			\mathsf{x} & \mathsf{x} & 0 \\
			\end{pmatrix}$
			\\[5mm]
			$\begin{pmatrix}
			0 & 0 & \mathsf{x} \\
			\mathsf{x} & \mathsf{x} & \mathsf{x} \\
			0 & \mathsf{x} & \mathsf{x} \\
			\end{pmatrix}$
			& $\begin{pmatrix}
			0 & 0 & \mathsf{x} \\
			\mathsf{x} & \mathsf{x} & \mathsf{x} \\
			\mathsf{x} & 0 & \mathsf{x} \\
			\end{pmatrix}$
			& $\begin{pmatrix}
			\mathsf{x} & 0 & 0 \\
			\mathsf{x} & \mathsf{x} & \mathsf{x} \\
			\mathsf{x} & \mathsf{x} & 0 \\
			\end{pmatrix}$
			& $\begin{pmatrix}
			0 & \mathsf{x} & 0 \\
			\mathsf{x} & \mathsf{x} & \mathsf{x} \\
			0 & \mathsf{x} & \mathsf{x} \\
			\end{pmatrix}$
			& $\begin{pmatrix}
			0 & \mathsf{x} & 0 \\
			\mathsf{x} & \mathsf{x} & \mathsf{x} \\
			\mathsf{x} & \mathsf{x} & 0 \\
			\end{pmatrix}$
			& $\begin{pmatrix}
			\mathsf{x} & 0 & 0 \\
			\mathsf{x} & \mathsf{x} & \mathsf{x} \\
			\mathsf{x} & 0 & \mathsf{x} \\
			\end{pmatrix}$
			& $\begin{pmatrix}
			0 & \mathsf{x} & \mathsf{x} \\
			\mathsf{x} & 0 & \mathsf{x} \\
			0 & \mathsf{x} & \mathsf{x} \\
			\end{pmatrix}$
			& $\begin{pmatrix}
			\mathsf{x} & 0 & \mathsf{x} \\
			0 & \mathsf{x} & \mathsf{x} \\
			\mathsf{x} & 0 & \mathsf{x} \\
			\end{pmatrix}$
			& $\begin{pmatrix}
			\mathsf{x} & \mathsf{x} & 0 \\
			\mathsf{x} & 0 & \mathsf{x} \\
			\mathsf{x} & \mathsf{x} & 0 \\
			\end{pmatrix}$
			\\[5mm]
			$\begin{pmatrix}
			0 & \mathsf{x} & \mathsf{x} \\
			\mathsf{x} & \mathsf{x} & 0 \\
			0 & \mathsf{x} & \mathsf{x} \\
			\end{pmatrix}$
			& $\begin{pmatrix}
			\mathsf{x} & \mathsf{x} & 0 \\
			0 & \mathsf{x} & \mathsf{x} \\
			\mathsf{x} & \mathsf{x} & 0 \\
			\end{pmatrix}$
			& $\begin{pmatrix}
			\mathsf{x} & 0 & \mathsf{x} \\
			\mathsf{x} & \mathsf{x} & 0 \\
			\mathsf{x} & 0 & \mathsf{x} \\
			\end{pmatrix}$
			& $\begin{pmatrix}
			0 & \mathsf{x} & \mathsf{x} \\
			\mathsf{x} & \mathsf{x} & \mathsf{x} \\
			0 & 0 & \mathsf{x} \\
			\end{pmatrix}$
			& $\begin{pmatrix}
			\mathsf{x} & 0 & \mathsf{x} \\
			\mathsf{x} & \mathsf{x} & \mathsf{x} \\
			0 & 0 & \mathsf{x} \\
			\end{pmatrix}$
			& $\begin{pmatrix}
			\mathsf{x} & \mathsf{x} & 0 \\
			\mathsf{x} & \mathsf{x} & \mathsf{x} \\
			\mathsf{x} & 0 & 0 \\
			\end{pmatrix}$
			& $\begin{pmatrix}
			0 & \mathsf{x} & \mathsf{x} \\
			\mathsf{x} & \mathsf{x} & \mathsf{x} \\
			0 & \mathsf{x} & 0 \\
			\end{pmatrix}$
			& $\begin{pmatrix}
			\mathsf{x} & \mathsf{x} & 0 \\
			\mathsf{x} & \mathsf{x} & \mathsf{x} \\
			0 & \mathsf{x} & 0 \\
			\end{pmatrix}$
			& $\begin{pmatrix}
			\mathsf{x} & 0 & \mathsf{x} \\
			\mathsf{x} & \mathsf{x} & \mathsf{x} \\
			\mathsf{x} & 0 & 0 \\
			\end{pmatrix}$
			\\[5mm]
			$\begin{pmatrix}
			\mathsf{x} & 0 & \mathsf{x} \\
			0 & \mathsf{x} & \mathsf{x} \\
			0 & \mathsf{x} & \mathsf{x} \\
			\end{pmatrix}$
			& $\begin{pmatrix}
			0 & \mathsf{x} & \mathsf{x} \\
			\mathsf{x} & 0 & \mathsf{x} \\
			\mathsf{x} & 0 & \mathsf{x} \\
			\end{pmatrix}$
			& $\begin{pmatrix}
			\mathsf{x} & 0 & \mathsf{x} \\
			\mathsf{x} & \mathsf{x} & 0 \\
			\mathsf{x} & \mathsf{x} & 0 \\
			\end{pmatrix}$
			& $\begin{pmatrix}
			\mathsf{x} & \mathsf{x} & 0 \\
			0 & \mathsf{x} & \mathsf{x} \\
			0 & \mathsf{x} & \mathsf{x} \\
			\end{pmatrix}$
			& $\begin{pmatrix}
			0 & \mathsf{x} & \mathsf{x} \\
			\mathsf{x} & \mathsf{x} & 0 \\
			\mathsf{x} & \mathsf{x} & 0 \\
			\end{pmatrix}$
			& $\begin{pmatrix}
			\mathsf{x} & \mathsf{x} & 0 \\
			\mathsf{x} & 0 & \mathsf{x} \\
			\mathsf{x} & 0 & \mathsf{x} \\
			\end{pmatrix}$
			& $\begin{pmatrix}
			\mathsf{x} & \mathsf{x} & \mathsf{x} \\
			0 & 0 & \mathsf{x} \\
			0 & \mathsf{x} & \mathsf{x} \\
			\end{pmatrix}$
			& $\begin{pmatrix}
			\mathsf{x} & \mathsf{x} & \mathsf{x} \\
			0 & 0 & \mathsf{x} \\
			\mathsf{x} & 0 & \mathsf{x} \\
			\end{pmatrix}$
			& $\begin{pmatrix}
			\mathsf{x} & \mathsf{x} & \mathsf{x} \\
			\mathsf{x} & 0 & 0 \\
			\mathsf{x} & \mathsf{x} & 0 \\
			\end{pmatrix}$
			\\[5mm]
			$\begin{pmatrix}
			\mathsf{x} & \mathsf{x} & \mathsf{x} \\
			0 & \mathsf{x} & 0 \\
			0 & \mathsf{x} & \mathsf{x} \\
			\end{pmatrix}$
			& $\begin{pmatrix}
			\mathsf{x} & \mathsf{x} & \mathsf{x} \\
			0 & \mathsf{x} & 0 \\
			\mathsf{x} & \mathsf{x} & 0 \\
			\end{pmatrix}$
			& $\begin{pmatrix}
			\mathsf{x} & \mathsf{x} & \mathsf{x} \\
			\mathsf{x} & 0 & 0 \\
			\mathsf{x} & 0 & \mathsf{x} \\
			\end{pmatrix}$
			& $\begin{pmatrix}
			\mathsf{x} & \mathsf{x} & \mathsf{x} \\
			0 & \mathsf{x} & \mathsf{x} \\
			0 & 0 & \mathsf{x} \\
			\end{pmatrix}$
			& $\begin{pmatrix}
			\mathsf{x} & \mathsf{x} & \mathsf{x} \\
			\mathsf{x} & 0 & \mathsf{x} \\
			0 & 0 & \mathsf{x} \\
			\end{pmatrix}$
			& $\begin{pmatrix}
			\mathsf{x} & \mathsf{x} & \mathsf{x} \\
			\mathsf{x} & \mathsf{x} & 0 \\
			\mathsf{x} & 0 & 0 \\
			\end{pmatrix}$
			& $\begin{pmatrix}
			\mathsf{x} & \mathsf{x} & \mathsf{x} \\
			0 & \mathsf{x} & \mathsf{x} \\
			0 & \mathsf{x} & 0 \\
			\end{pmatrix}$
			& $\begin{pmatrix}
			\mathsf{x} & \mathsf{x} & \mathsf{x} \\
			\mathsf{x} & \mathsf{x} & 0 \\
			0 & \mathsf{x} & 0 \\
			\end{pmatrix}$
			& $\begin{pmatrix}
			\mathsf{x} & \mathsf{x} & \mathsf{x} \\
			\mathsf{x} & 0 & \mathsf{x} \\
			\mathsf{x} & 0 & 0 \\
			\end{pmatrix}$
			\\[5mm]
			\hline
		\end{tabular*}
\end{table*}

Our first goal is to obtain the maximal number of zeroes in an arbitrary quark mass matrix by applying only RhWBT. Since  such transformations can be applied individually to each quark sector, we can start our analysis from  a general $3\times3$ complex mass matrix
\begin{equation}
M=\begin{pmatrix}
 \mathsf{x} & \mathsf{x} & \mathsf{x} \\
 \mathsf{x} & \mathsf{x} & \mathsf{x} \\ 
 \mathsf{x} & \mathsf{x} & \mathsf{x} 
\end{pmatrix},
\end{equation}
where $\mathsf{x}$ stands for a nonvanishing element. 

From the mathematical point of view, a unitary matrix is a composition of the rotation matrices $O_{ij}(\theta)$, defined as
\begin{subequations}
\begin{equation}
O_{12}(\theta) = 
\begin{pmatrix}
 \cos\theta & \sin\theta & 0\\
 -\sin\theta & \cos\theta & 0\\
  0 & 0 & 1
\end{pmatrix},
\end{equation}
\begin{equation}
O_{13}(\theta) =
\begin{pmatrix}
 \cos\theta & 0 & \sin\theta\\
  0 & 1 & 0 \\ 
  -\sin\theta & 0 & \cos\theta
\end{pmatrix},
\end{equation}
\begin{equation}
O_{23}(\theta) =
\begin{pmatrix}
 1 & 0 & 0\\
  0 & \cos\theta & \sin\theta\\
  0 &-\sin\theta & \cos\theta
\end{pmatrix},
\end{equation}
\end{subequations}
combined with arbitrary phase diagonal matrices. Other useful WB transformations that can be invoked are permutations, which however do not change the number of zeroes. The six permutation matrices are
\begin{equation}
\begin{aligned}
&P_e =
\left(\begin{array}{@{}c@{\;}c@{\;}c@{}}
1 & 0 & 0 \\ 0 & 1 & 0\\ 0 & 0 & 1
\end{array}\right), \,
P_{(12)} =
\left(\begin{array}{@{}c@{\;}c@{\;}c@{}}
0 & 1 & 0\\ 1 & 0 & 0\\ 0 & 0 & 1
\end{array}\right), \,
P_{(13)} =
\left(\begin{array}{@{}c@{\;}c@{\;}c@{}}
0 & 0 & 1\\ 0 & 1 & 0\\ 1 & 0 & 0
\end{array}\right),  \\
&P_{(23)} =
\left(\begin{array}{@{}c@{\;}c@{\;}c@{}}
1 & 0 & 0\\ 0 & 0 & 1\\ 0 & 1 & 0
\end{array}\right), \,
P_{(123)} =
\left(\begin{array}{@{}c@{\;}c@{\;}c@{}}
0 & 0 & 1\\ 1 & 0 & 0\\ 0 & 1 & 0
\end{array}\right), \,
P_{(132)} =
\left(\begin{array}{@{}c@{\;}c@{\;}c@{}}
0 & 1 & 0\\ 0 & 0 & 1\\ 1 & 0 & 0
\end{array}\right).
\end{aligned}
\end{equation}

Obtaining the first zero is trivial, one just needs a single rotation. For instance, in order to have a zero element at the position $(1,1)$ of $M$, we apply the transformation $M'=MKO_{12}(\theta)$, with $K=\diag(1,e^{-i\varphi},1)$, so that
\begin{equation}
M'_{11}=M_{11}\cos\theta-M_{12}e^{-i\varphi}\sin\theta.
\end{equation}
Choosing $\tan\theta=|M_{11}|/|M_{12}|$ and $\varphi=\arg(M^{\phantom{\ast}}_{11}M^{\ast}_{12})$ the desired zero at position $(1,1)$ is obtained,
\begin{equation}
\label{eq:Mp}
M'=\begin{pmatrix}
 0 & \mathsf{x} & \mathsf{x} \\
 \mathsf{x} & \mathsf{x} & \mathsf{x} \\ 
 \mathsf{x} & \mathsf{x} & \mathsf{x} 
\end{pmatrix}.
\end{equation}
Following a similar procedure, and using a suitable right-handed rotation, one can conclude that any element of the mass matrix $M$ can be enforced to be zero.

The next question to address is whether a second zero can be obtained in the matrix $M'$. Adding a new zero in a different column is straightforward; we simply apply a new right-handed rotation to $M'$ in the sector that does not involve the zero element of this matrix. Consider, for instance, the matrix in Eq.~\eqref{eq:Mp}. Acting with $O_{23}(\theta)$ on the right of this matrix, in combination with the appropriate diagonal phase matrix $K'$, 
\begin{equation}
\label{eq:Mpp}
M''=M'K'O_{23}(\theta),
\end{equation}
a new zero can be created at any matrix element of the second and third columns of $M''$. 

\begin{table*}[t]
	\caption{\label{tab:3NWB} The 24 nonsingular mass matrices $M$ with three zeros that cannot be obtained through RhWBT. The matrices are classified into four classes according to the corresponding hermitian matrix $H=M M^{\dagger}$.}
		\begin{tabular*}{\linewidth}{@{\extracolsep{\fill}}ccccccc}
			\hline
			Class & \multicolumn{6}{c}{Textures of matrix $M$}\\
			\hline
			$H_{ij}\neq0$ & $\begin{pmatrix}
			0 & \mathsf{x} & \mathsf{x} \\
			\mathsf{x} & 0 & \mathsf{x} \\
			\mathsf{x} & \mathsf{x} & 0 \\
			\end{pmatrix}$
			& $\begin{pmatrix}
			0 & \mathsf{x} & \mathsf{x} \\
			\mathsf{x} & \mathsf{x} & 0 \\
			\mathsf{x} & 0 & \mathsf{x} \\
			\end{pmatrix}$
			& $\begin{pmatrix}
			\mathsf{x} & 0 & \mathsf{x} \\
			0 & \mathsf{x} & \mathsf{x} \\
			\mathsf{x} & \mathsf{x} & 0 \\
			\end{pmatrix}$
			& $\begin{pmatrix}
			\mathsf{x} & 0 & \mathsf{x} \\
			\mathsf{x} & \mathsf{x} & 0 \\
			0 & \mathsf{x} & \mathsf{x} \\
			\end{pmatrix}$
			& $\begin{pmatrix}
			\mathsf{x} & \mathsf{x} & 0 \\
			0 & \mathsf{x} & \mathsf{x} \\
			\mathsf{x} & 0 & \mathsf{x} \\
			\end{pmatrix}$
			& $\begin{pmatrix}
			\mathsf{x} & \mathsf{x} & 0 \\
			\mathsf{x} & 0 & \mathsf{x} \\
			0 & \mathsf{x} & \mathsf{x} \\
			\end{pmatrix}$
			\\[5mm]
			$H_{12}=0$ & $\begin{pmatrix}
			0 & 0 & \mathsf{x} \\
			\mathsf{x} & \mathsf{x} & 0 \\
			\mathsf{x} & \mathsf{x} & \mathsf{x} \\
			\end{pmatrix}$
			& $\begin{pmatrix}
			0 & \mathsf{x} & 0 \\
			\mathsf{x} & 0 & \mathsf{x} \\
			\mathsf{x} & \mathsf{x} & \mathsf{x} \\
			\end{pmatrix}$
			& $\begin{pmatrix}
			0 & \mathsf{x} & \mathsf{x} \\
			\mathsf{x} & 0 & 0 \\
			\mathsf{x} & \mathsf{x} & \mathsf{x} \\
			\end{pmatrix}$
			& $\begin{pmatrix}
			\mathsf{x} & 0 & 0 \\
			0 & \mathsf{x} & \mathsf{x} \\
			\mathsf{x} & \mathsf{x} & \mathsf{x} \\
			\end{pmatrix}$
			& $\begin{pmatrix}
			\mathsf{x} & 0 & \mathsf{x} \\
			0 & \mathsf{x} & 0 \\
			\mathsf{x} & \mathsf{x} & \mathsf{x} \\
			\end{pmatrix}$
			& $\begin{pmatrix}
			\mathsf{x} & \mathsf{x} & 0 \\
			0 & 0 & \mathsf{x} \\
			\mathsf{x} & \mathsf{x} & \mathsf{x} \\
			\end{pmatrix}$
			\\[5mm]
			$H_{13}=0$ &$\begin{pmatrix}
			0 & 0 & \mathsf{x} \\
			\mathsf{x} & \mathsf{x} & \mathsf{x} \\
			\mathsf{x} & \mathsf{x} & 0 \\
			\end{pmatrix}$
			& $\begin{pmatrix}
			0 & \mathsf{x} & 0 \\
			\mathsf{x} & \mathsf{x} & \mathsf{x} \\
			\mathsf{x} & 0 & \mathsf{x} \\
			\end{pmatrix}$
			& $\begin{pmatrix}
			0 & \mathsf{x} & \mathsf{x} \\
			\mathsf{x} & \mathsf{x} & \mathsf{x} \\
			\mathsf{x} & 0 & 0 \\
			\end{pmatrix}$
			& $\begin{pmatrix}
			\mathsf{x} & 0 & 0 \\
			\mathsf{x} & \mathsf{x} & \mathsf{x} \\
			0 & \mathsf{x} & \mathsf{x} \\
			\end{pmatrix}$
			& $\begin{pmatrix}
			\mathsf{x} & 0 & \mathsf{x} \\
			\mathsf{x} & \mathsf{x} & \mathsf{x} \\
			0 & \mathsf{x} & 0 \\
			\end{pmatrix}$
			& $\begin{pmatrix}
			\mathsf{x} & \mathsf{x} & 0 \\
			\mathsf{x} & \mathsf{x} & \mathsf{x} \\
			0 & 0 & \mathsf{x} \\
			\end{pmatrix}$
			\\[5mm]
			$H_{23}=0$ &$\begin{pmatrix}
			\mathsf{x} & \mathsf{x} & \mathsf{x} \\
			0 & 0 & \mathsf{x} \\
			\mathsf{x} & \mathsf{x} & 0 \\
			\end{pmatrix}$
			& $\begin{pmatrix}
			\mathsf{x} & \mathsf{x} & \mathsf{x} \\
			0 & \mathsf{x} & 0 \\
			\mathsf{x} & 0 & \mathsf{x} \\
			\end{pmatrix}$
			& $\begin{pmatrix}
			\mathsf{x} & \mathsf{x} & \mathsf{x} \\
			0 & \mathsf{x} & \mathsf{x} \\
			\mathsf{x} & 0 & 0 \\
			\end{pmatrix}$
			& $\begin{pmatrix}
			\mathsf{x} & \mathsf{x} & \mathsf{x} \\
			\mathsf{x} & 0 & 0 \\
			0 & \mathsf{x} & \mathsf{x} \\
			\end{pmatrix}$
			& $\begin{pmatrix}
			\mathsf{x} & \mathsf{x} & \mathsf{x} \\
			\mathsf{x} & 0 & \mathsf{x} \\
			0 & \mathsf{x} & 0 \\
			\end{pmatrix}$
			& $\begin{pmatrix}
			\mathsf{x} & \mathsf{x} & \mathsf{x} \\
			\mathsf{x} & \mathsf{x} & 0 \\
			0 & 0 & \mathsf{x} \\
			\end{pmatrix}$
			\\[5mm]
			\hline
		\end{tabular*}
\end{table*}

One may also ask whether it is possible to have simultaneously two zeroes in the same column by means of right-handed unitary transformations. In order to see that this is indeed the case, let us start with a matrix $M''$, obtained via Eq.~\eqref{eq:Mpp}, where the second zero is at the position $(2,2)$, i.e.,
\begin{equation}
\label{eq:Mpp22}
M''=\begin{pmatrix}
 0 & \mathsf{x} & \mathsf{x} \\
 \mathsf{x} & 0 & \mathsf{x} \\ 
 \mathsf{x} & \mathsf{x} & \mathsf{x} 
\end{pmatrix}.
\end{equation}
First we note that, without loss of generality, the matrix elements $M''_{12}$, $M''_{13}$, $M''_{21}$, and $M''_{23}$ can be assumed real, since they do not form a quartet. Now we perform the following right-handed unitary transformations:
\begin{equation}
\widetilde{M}''=M''\,O_{12}(\theta_1)\,O_{13}(\theta_2),
\end{equation}
so that the matrix elements 
\begin{subequations}
\begin{align}
\widetilde{M}''_{11}&=-M''_{12}\sin\theta_1\cos\theta_2-M''_{13}\sin\theta_2,\\
\widetilde{M}''_{21}&=M''_{21}\cos\theta_1\cos\theta_2  - M''_{23}\sin\theta_2,
\end{align}
\end{subequations}
vanish, i.e., $\widetilde{M}''_{11}=\widetilde{M}''_{21}=0$. The solutions of these equations are then
\begin{equation}
\tan\theta_1 = -\frac{M''_{21}M''_{13}}{M''_{12}M''_{23}},\quad
\tan\theta_2 = -\frac{M''_{12}}{M''_{13}}\sin\theta_1,
\end{equation}
leading to the desired pattern
\begin{equation}
\label{eq:Mpptilde}
\widetilde{M}''=\begin{pmatrix}
 0 & \mathsf{x} & \mathsf{x} \\
 0 & \mathsf{x} & \mathsf{x} \\ 
 \mathsf{x} & \mathsf{x} & \mathsf{x} 
\end{pmatrix}.
\end{equation}
We conclude that there are 36 textures with two zeroes, which  are all equivalent in the sense that they can be obtained by means of right-handed unitary transformations.

Let us now consider the possibility of adding a third zero through a RhWBT. There exist 84 textures with three zeroes. Out of them, 6 matrices have determinant zero (a null row or a null column). The remaining 78 nonsingular matrices are distributed in four classes, depending on the texture of the hermitian matrix $H=M M^{\dagger}$. There are 60 matrices for which $H$ has no zeroes, 6 matrices with $H_{12}=0$, 6 matrices with $H_{13}=0$, and 6 matrices with $H_{23}=0$. In the latter three classes, matrices within the same class are equivalent under RhWBT.  From the class with nonvanishing $H$, two subclasses can be distinguished. The first subclass includes 6 matrices, namely, the matrix with null diagonal and those obtained by its column permutations, which by themselves are not weak bases. On the other hand, the 54 matrices of the second subclass (see Table~\ref{tab:3ZWB}) are all RhWB equivalent. The remaining 24 nonsingular mass matrices with three zeros, classified according to the classes of the corresponding matrix $H$, cannot be obtained by means of RhWBT. These matrices are listed in Table~\ref{tab:3NWB}.

Constructing a third WB zero is then an easy task. Starting from a matrix with two WB zeros in the same column, we apply a new right-handed rotation to this matrix in a sector that does not involve the column with zeros. For instance, starting from the matrix in Eq.~\eqref{eq:Mpptilde} and applying a rotation $O_{23}(\theta)$ on the right of this matrix, the third zero can be created at any matrix element of the second and third columns of $\widetilde{M}''$. Following this procedure, the 54 matrices given in Table~\ref{tab:3ZWB} can be obtained. 

As an example, let us construct the fourth matrix in the second row of Table~\ref{tab:3ZWB}, which is the first texture in the table that has no horizontally aligned zeroes. We act with $O_{23}(\theta)$ on the right of the matrix $\widetilde{M}''$ in Eq.~\eqref{eq:Mpptilde}, in combination with the diagonal phase matrix  $K''=\diag(1,1,e^{-i\varphi})$, so that $M'''=\widetilde{M}''K''O_{23}(\theta)$, with $M_{32}'''=0$. The latter condition is verified if
\begin{equation}
M_{32}'''=\widetilde{M}''_{32}\cos\theta-\widetilde{M}''_{33}e^{-i\varphi}\sin\theta=0.
\end{equation}
Choosing $\tan\theta=|\widetilde{M}''_{32}|/|\widetilde{M}''_{33}|$ and $\varphi=\arg(\widetilde{M}''^{\phantom{\ast}}_{33} {\widetilde{M}''^{\ast}_{32}})$, we then arrive at the zero texture form
\begin{equation}
\label{eq:Mppp}
M'''=\begin{pmatrix}
0 & \mathsf{x} & \mathsf{x} \\
0 & \mathsf{x} & \mathsf{x} \\ 
\mathsf{x} & 0 & \mathsf{x} 
\end{pmatrix}.
\end{equation}

Before concluding this section, let us briefly comment on the mass matrix patterns with more than three zeros (see also Table~\ref{tab:classes}). First we note that none of the textures given in Table~\ref{tab:3ZWB} can give rise to four-zero WB textures by only performing RhWBT. This is due to the fact that it is not possible to add a fourth zero in these textures without destroying one of the existing zeroes. The same argument applies to the 6 textures given in the first row of Table~\ref{tab:3NWB}. As for the remaining 18 textures given in Table~\ref{tab:3NWB}, namely, those that contain only one vanishing off-diagonal matrix element in $H$, it is easy to conclude that one can always add a fourth zero by means of a right-handed rotation. However, none of such four-zero textures corresponds to a weak basis. In a similar way, one can conclude that none of the textures with five or six zeroes can be obtained through RhWBT.

\section{General weak basis transformations}
\label{sec:genWB}

Next we consider the possibility of applying general WB transformations in order to obtain other WB pairs of quark mass matrices with the maximal number of texture zeroes. In particular, we shall look for left-handed and right-handed rotations, which, after being applied to the quark mass matrices, lead to pairs  with five and four WB texture zeroes.

Let us start from a WB where the up-quark mass matrix has six zeroes (it could be the diagonal matrix or any of its row and column permutations). As shown in the previous section, through right-handed rotations we can obtain 54 down-quark mass matrices with three zeroes. Clearly, this set of mass matrix pairs $M_u$, $M_d$ are all WB equivalent and thus do not have physical content. Examining the textures in Table~\ref{tab:3ZWB}, one concludes that all matrices have only one quartet without vanishing elements, which is directly related to the CP-odd WB invariant as
\begin{multline}
\im\Tr\left[
M^{\phantom{\dagger}}_uM^{\dagger}_u,M^{\phantom{\dagger}}_dM^{\dagger}_d
\right]^3\propto\\
\sin\left\{\arg\left[(M_d)^{\phantom{\ast}}_{ij}(M_d)^{\phantom{\ast}}_{kl}(M_d)^{\ast}_{il}(M_d)^{\ast}_{kj}\right]\right\},
\end{multline}
where $i$, $j$, $k$ and $l$ are the indices of the quartet elements.

Using any of these bases, it is easy to construct a new WB basis in which the matrix $M_u$ has 5 zeroes and $M_d$ has 4 zeroes (or vice versa). In other to accomplish this, one first observes that $M_d$ has necessarily one column with two zeroes (see Table~\ref{tab:3ZWB}). One can then multiply both quark mass matrices by a single left-handed rotation that maintains intact the three zeroes in $M_d$ and creates the fourth zero. This is only possible when the third zero in $M_d$ is not horizontally aligned with one of the two zeroes in the same column. As can be seen from Table~\ref{tab:3ZWB}, there are only 18 possibilities. Once the left-handed rotation is performed, the matrix $M_u$ becomes decoupled (i.e., with two zeroes in one column and two zeroes in one row). One can then perform a right-handed rotation on $M_u$ that maintains the two zeroes in the row and simultaneously adds the desired fifth zero. Notice that there is still freedom to apply permutation matrices on the right and on the left of both quark mass matrices (provided that the same left-handed permutation is applied in both sectors) in order to change the positions of the zeroes and generate new pairs of textures. Furthermore, in some special cases, it is also possible to change the positions of certain zeros by applying only right-handed rotations. Through these paths, we end up with 2592 different pairs of nonsingular mass matrices $M_u$ and $M_d$ that contain five and four zeroes, respectively. All these pairs are WB choices. The remaining 324 pairs have necessarily a decoupled  down-quark sector and, therefore, are neither weak bases nor viable ans\"{a}tze.

In order to illustrate the above procedure, next we consider the well-known NNI quark texture~\cite{Branco:1988iq,Harayama:1996am,Koide:1997ai,Branco:2010tx}, described by the pattern
\begin{equation}
\begin{pmatrix}
0 & \mathsf{x} & 0 \\
\mathsf{x} & 0 &  \mathsf{x}\\
0 & \mathsf{x} & \mathsf{x} 
\end{pmatrix}.
\end{equation}
More precisely, we construct WB pairs with $M_u$ and $M_d$ having the NNI form with five and four zeroes, respectively. Such pairs contain only 10 parameters (to be compared with the usual 12-parameter NNI form with 4 zeroes in each quark sector~\cite{Branco:1988iq}). Notice that the fifth zero in $M_u$ can only be located in the matrix elements $(2,3)$ and $(3,2)$; otherwise, the NNI form would have a null determinant~\cite{Harayama:1996am,Koide:1997ai}.

Let us start, for instance, from the weak basis in which the up-quark mass matrix is real diagonal and the down-quark mass matrix has the form given in Eq.~\eqref{eq:Mppp}, i.e.,
\begin{equation}
M_u=\begin{pmatrix}
 \mathsf{x} & 0 & 0 \\
0 & \mathsf{x} & 0 \\ 
0 & 0 & \mathsf{x} 
\end{pmatrix},\quad
M_d=\begin{pmatrix}
0 & \mathsf{x} & \mathsf{x} \\
0 & \mathsf{x} & \mathsf{x} \\ 
\mathsf{x} & 0 & \mathsf{x} 
\end{pmatrix}.
\end{equation}
Without loss of generality, we assume ${(M_d)}_{22}$ to be complex, while the remaining matrix elements in $M_d$ are real. We then apply the following WBT:
\begin{subequations}
\label{eq:nni}
\begin{align}
M'_u&=P^{\intercal}_{(23)}O^{\intercal}_{12}(\theta_1)\,M_u\,O_{12}(\theta_2)P_{(13)},\\
M'_d&=P^{\intercal}_{(23)}O^{\intercal}_{12}(\theta_1)\,M_d,
\end{align}
\end{subequations}
where 
\begin{equation}
\label{eq:nni-theta}
\tan\theta_1=\frac{{(M_d)}_{13}}{{(M_d)}_{23}},\quad
\tan\theta_2=\frac{{(M_u)}_{11}\,{(M_d)}_{23}}{{(M_u)}_{22}\,{(M_d)}_{13}}.
\end{equation}

Once the above WBT are carried out, we obtain the zero textures
\begin{equation} \label{NNI1}
M'_u=\begin{pmatrix}
0 & \mathsf{x} & 0 \\
\mathsf{x} & 0 & 0 \\
0 & \mathsf{x} & \mathsf{x} 
\end{pmatrix},\quad
M'_d=\begin{pmatrix}
0 & \mathsf{x} & 0 \\
\mathsf{x} & 0 &  \mathsf{x}\\
0 & \mathsf{x} & \mathsf{x} 
\end{pmatrix},
\end{equation}
which is precisely an NNI WB pair with an additional zero at the position (2,3) of $M'_u$. 

Let us now construct a second NNI pair in such a way that the additional zero is at the position (3,2) of $M_u$. First we remark that, in the process of constructing the matrix pair~\eqref{NNI1},  only the elements ${(M'_d)}_{12}$ and ${(M'_d)}_{32}$ become complex. Acting on this pair, it is now possible to perform the following set of WBT 
\begin{subequations}
\begin{align}
M''_u&=P^{\intercal}_{(12)}\,M'_u\,O_{23}(\theta'_1)P_{(12)},\\
M''_d&=P^{\intercal}_{(12)}\,M'_d\,O_{13}(\theta'_2)\,K\,O_{12}(\theta'_3)P_{(123)},
\end{align}
\end{subequations}
where $K=\diag(1,e^{-i\varphi},1)$, $\varphi=\arg{(M'_d)}_{32}$, and
\begin{equation}
\begin{aligned}
&\tan\theta'_1=\frac{(M'_u)_{32}}{{(M'_u)}_{33}},\quad
\tan\theta'_2=\frac{{(M'_d)}_{21}}{{(M'_d)}_{23}},\\[2mm]
&\tan\theta'_3=\frac{\left|{(M'_d)}_{32}\right|\sqrt{(M'_d)^2_{21}+(M'_d)^2_{23}}}{{(M'_d)}_{21}{(M'_d)}_{33}}.
\end{aligned}
\end{equation}
Once this is done, the new NNI pair 
\begin{equation}\label{NNI2}
M''_u=\begin{pmatrix}
0 & \mathsf{x} & 0 \\
\mathsf{x} & 0 & \mathsf{x} \\
0 & 0 & \mathsf{x} 
\end{pmatrix},\quad
M''_d=\begin{pmatrix}
0 & \mathsf{x} & 0 \\
\mathsf{x} & 0 &  \mathsf{x}\\
0 & \mathsf{x} & \mathsf{x} 
\end{pmatrix},
\end{equation}
is obtained.
As in the case of Eq.~\eqref{NNI1}, this WB pair contains less parameters than the usual NNI case with 4 zeroes in each quark sector.

\section{Numerical analysis}
\label{sec:numex}

In the previous sections, we have shown that in the basis where one quark mass matrix is diagonal, there are 54 three-zero textures in the other quark sector that simply correspond to WB choices and thus are all equivalent. This remarkable fact has been overlooked in previous works. For instance, in Ref.~\cite{Tanimoto:2016rqy}, in the framework of the so-called Occam's razor approach, three zeros in the matrix $M_d$ were imposed, assuming $M_u$ diagonal. Among the textures considered by the authors, the ones that are consistent with the experimental data precisely belong to the set of matrices listed in Table~\ref{tab:3ZWB}. Clearly, these WB pairs of matrices can always reproduce the ten physical observables and, without any further assumptions, they do not lead to physical predictions. Therefore, it is not surprising that the Cabibbo angle, as well as other CKM mixing angles and the CP-violating phase, are successfully predicted by such quark mass matrices.

Let us now analyse the 24 quark mass matrices listed in Table~\ref{tab:3NWB}. These matrices cannot be obtained through WB transformations. Assuming that $M_u$ (or $M_d$) is diagonal, one may then ask whether these three-zero textures are compatible with current experimental data. To test these ans\"atze, we shall employ in our numerical analysis a standard $\chi^2$ minimisation with respect to the ten observable quantities. The CKM parameters used in our analysis are the real quantities $\lambda$, $A$, $\overline{\rho}$ and $\overline{\eta}$, defined  in terms of rephasing invariant quantities as~\cite{Agashe:2014kda}
\begin{subequations}
\begin{align}
&\lambda\equiv\frac{|V_{us}|}{\sqrt{|V_{us}|^2+|V_{ud}|^2}},\\[2mm]
&A\equiv\frac1{\lambda}\left|\frac{V_{cb}}{V_{us}}\right|,\\[2mm]
&\overline{\rho}+i\,\overline{\eta} \equiv -\frac{V^{\phantom{\ast}}_{ud}V^{\ast}_{ub}}{V^{\phantom{\ast}}_{cd}V^{\ast}_{cb}},
\end{align}
\end{subequations}
where $V_{ij}$ are the usual CKM matrix elements. The quark masses at the electroweak scale are obtained using the input parameters given in Table~\ref{tab:inputs} and applying the QCD renormalisation group equations at four-loop level. The best-fit values of the physical observables and their $1\sigma$ errors are given in Table~\ref{tab:massMZ}.

\begin{table}[t]
	\caption{\label{tab:inputs} Input parameters used in our analysis to determine the quark mass values at the electroweak scale $M_Z=91.1876\pm0.0021\,\text{GeV}$. Pole masses are denoted by $M$, while $m_q(\mu)$ corresponds to the running masses in the $\overline{MS}$ scheme. Note that $\overline{m} \equiv (m_u+m_d)/2$ and $\alpha_s$ is the strong coupling constant.	}
	{\def\arraystretch{1.3}
		\begin{tabular*}{\linewidth}{@{\extracolsep{\fill}}ll}
			\hline
			\multicolumn{2}{c}{Inputs~\cite{Agashe:2014kda}}
			\\\hline
			$m_u(2\,\text{GeV})=2.3^{+0.7}_{-0.5}\,\text{MeV}$
			&
			$0.38< m_u/m_d<0.58$
			\\
			$m_d(2\,\text{GeV})=4.8^{+0.5}_{-0.3}\,\text{MeV}$
			&
			$17< m_s/m_d<22$
			\\
			$m_s(2\,\text{GeV})=95\pm5\,\text{MeV}$
			&
			$\overline{m}=3.5^{+0.7}_{-0.2}\,\text{MeV}$
			\\
			$m_c(m_c)=1.275\pm0.025\,\text{GeV}$
			&
			$m_s/\,\overline{m}=27.5\pm1.0$
			\\
			$m_b(m_b)=4.18\pm0.03\,\text{GeV}$
			&
			$\alpha_s(M_Z)=0.1181\pm0.0013$
			\\
			$M_t=174.6\pm1.9\,\text{GeV}$
			&
			\\
			\hline
		\end{tabular*}}
	\end{table}

\begin{table}[t]
\caption{\label{tab:massMZ}Quark masses, calculated at four-loop level, and CKM mixing parameters used in the numerical $\chi^2$-analysis.}
{\def\arraystretch{1.3}
\begin{tabular*}{\linewidth}{@{\extracolsep{\fill}}ll}
\hline
\multicolumn{1}{c}{Quark masses} &
\multicolumn{1}{c}{CKM parameters~\cite{Charles:2015gya}}\\
\hline
$m_u(M_Z)=1.327\pm0.28\,\text{MeV}$ & $\lambda=0.22548^{+0.00068}_{-0.00034}$\\
$m_d(M_Z)=2.769^{+0.33}_{-0.21}\,\text{MeV}$ & $A=0.810^{+0.018}_{-0.024}$\\
$m_s(M_Z)=54.79\pm3.6\,\text{MeV}$ & $\overline{\rho}= 0.145^{+0.013}_{-0.007}$\\
$m_c(M_Z)=0.6314\pm0.031\,\text{GeV}$ & $\overline{\eta}=0.343^{+0.011}_{-0.012}$\\
$m_b(M_Z)=2.861\pm0.045\,\text{GeV}$\\
$m_t(M_Z)=173.0\pm2.1\,\text{GeV}$\\
\hline
\end{tabular*}}
\end{table}

We have performed a $\chi^2$-analysis using the standard definition of the $\chi^2$-function in terms of the observable values and their deviations from the mean values. The results are presented in Table~\ref{tab:3NWBchi2}. As can be seen from the table, none of the 24 pairs of textures is compatible with the experimental data. For the last three classes in the table, this result should not come as a surprise. Indeed, all these sets conserve the CP symmetry since, in the basis where a quark mass matrix is diagonal, the CP-odd WB invariant that controls  CP violation is proportional to $\im\left[H^{\phantom{\ast}}_{12}H^{\phantom{\ast}}_{23}H^{\ast}_{13}\right]$, and thus vanishes.

\begin{table}[t]
	\caption{\label{tab:3NWBchi2} The minimum of $\chi^2$ for the mass matrices given in Table~\ref{tab:3NWB}. The cases of $M_u=D_u$ and $M_d=D_d$ are considered.}
	{\def\arraystretch{1.3}\tabcolsep=12pt
		\begin{tabular*}{\linewidth}{@{\extracolsep{\fill}}ccc}
			\hline
			Class & $\chi^2_\text{min}(M_u=D_u)$ & $\chi^2_\text{min}(M_d=D_d)$\\
			\hline
			$H_{ij} \neq 0$ & $7.77\times10^2$ & $1.14\times10^3$\\
			$H_{12}=0$ & $1.43\times10^3$ & $5.20\times10^3$\\
			$H_{13}=0$ & $1.24\times10^3$ & $5.14\times10^3$\\
			$H_{23}=0$ & $9.79\times10^4$ & $1.54\times10^5$\\
			\hline
		\end{tabular*}}
\end{table}

\section{Conclusions}
\label{sec:summary}

In this work, we have searched for all weak bases that lead to texture zeroes in the quark mass matrices and contain a minimal number of parameters. We have shown that the maximum number of texture zeroes than can be obtained through WB transformations in both quark sectors is altogether nine. In fact, this number also corresponds to the maximum number of viable texture zeroes for a given quark mass matrix pair $M_u$ and $M_d$. In other words, there is no mass matrix combination with ten or more zeroes that is compatible with the experimental data. This conclusion is independent of the origin of the texture zeroes. We also remark that our results remain valid for the leptonic sector, provided that all light neutrino masses are distinct from zero and neutrinos are of Dirac nature.

Clearly, any WB pair of quark mass matrices can explain the ten physical observables (i.e., six nonvanishing quark masses, three mixing angles and one CP-violating phase), since these sets by themselves do not have physical implications. The nine zero entries can only be distributed between the up- and down-quark sectors in matrix pairs with six and three texture zeroes, or five and four texture zeroes. We have found that in the WB where a quark mass matrix is nonsingular and has six zeroes in one sector, there are 54 matrices with three zeroes in the other sector that can be obtained applying only right-handed WB transformations. We have also revealed that all pairs composed of a nonsingular matrix with five zeroes and a nonsingular and nondecoupled matrix with four zeroes simply correspond to WB choices. Once again, none of these pairs of quark mass matrices has physical content. We have also shown how to construct the well-known NNI WB patterns, starting from an arbitrary weak basis.

As a final comment, it is worthwhile to remark that our analysis has been performed at the electroweak scale. Nevertheless, all the WB texture zeroes are clearly stable under renormalisation group effects, since at any given energy scale one can always perform a WB transformation in order to preserve the position of the zeroes.

\section*{Acknowledgments}
The work of D.E.C. is supported by Associa\c c\~ ao do Instituto Superior T\'ecnico para a Investiga\c c\~ao e Desenvolvimento (IST-ID). The authors acknowledge support from Funda\c c\~ ao para a Ci\^ encia e a Tecnologia (FCT, Portugal) through the projects UID/FIS/00777/2013 and CERN/FIS-NUC/0010/2015. 


\bibliography{refs}

\end{document}